\documentclass[conference]{IEEEtran}
\IEEEoverridecommandlockouts
\usepackage{cite}
\usepackage{amsmath,amssymb,amsfonts}
\usepackage{algorithmic}
\usepackage{graphicx}
\usepackage{graphics}
\usepackage{textcomp}
\usepackage{tabularx}
\usepackage{xcolor}
\usepackage{hyphenat}
\usepackage{hyperref}
\def\BibTeX{{\rm B\kern-.05em{\sc i\kern-.025em b}\kern-.08em
    T\kern-.1667em\lower.7ex\hbox{E}\kern-.125emX}}
\begin{document}

\title{Performance Analysis of Zero-Trust multi-cloud\\
}
\author{
    \IEEEauthorblockN{Simone Rodigari, Donna O'Shea, 
    	Pat McCarthy,  Martin McCarry\IEEEauthorrefmark{1}, Sean McSweeney}
    \IEEEauthorblockA{
    Munster Technological University, Cork, Ireland
    }
    {\IEEEauthorrefmark{1} Pilz Ireland Industrial Automation Limited, Cork, Ireland}
    \IEEEauthorblockA{
    \{1\}{simone.rodigari}@mycit.ie, 
    \{2, 3, 5\}{firstname.lastname}@cit.ie, 
    \{4\}\IEEEauthorrefmark{1}{m.mccarry}@pilz.ie}

}

\maketitle

\begin{abstract}
Zero Trust security model permits to secure cloud native applications while encrypting all network communication, authenticating, and authorizing every request. The service mesh can enable Zero Trust using a side-car proxy without changes to the application code. To the best of our knowledge, no previous work has provided a performance analysis of Zero Trust in a multi-cloud environment. This paper proposes a multi-cloud framework and a testing workflow to analyse performance of the data plane under load and the impact on the control plane, when Zero Trust is enabled. The results of preliminary tests show that Istio has reduced latency variability in responding to sequential HTTP requests. Results also reveal that the overall CPU and memory usage can increase based on service mesh configuration and the cloud environment.
\end{abstract}


\section{Introduction}


Open-source solutions including Kubernetes, Apache Mesos and Docker Swarm\cite{al2019container}, provide microservices run-time environments and offer the necessary networking infrastructure for containerized microservices to communicate with each other and external clients on the public network\cite{b4}. Microservices in Kubernetes can run over multi clouds with multi-tenanted architecture, sharing compute resources with other untrusted applications. Traditionally, organizations configured trust environments to implement network security; users and services with access to the environment were considered trusted and allowed to execute operations within the system. This security model can also be defined as “Trustworthy” and is suitable for monolith applications, it typically involves the configuration of a firewall which provides the environment perimeter.
“Zero Trust” is a recent security model that does not have the concept of environment perimeter and it is more suitable for cloud applications. In a Zero Trust network the authentication, authorization and encryption are enforced on every communication between users – device – applications. The "Zero Trust" model addresses some of the challenges involved in a microservices deployments across heterogeneous networks including consistent and standardized enforcement of security policies. 

In this paper we evaluate the performance, in terms of latency and physical resources (CPU, memory) of a Zero Trust implementation using Istio \cite{istio} service mesh in a multi-cloud system composed of two separate Public Cloud Providers(PCP) to determine if there is a performance penalty at the data-plane level when enabling Zero Trust with a side-car proxy implementation in a multi-cloud architecture.
\section{Background}
\label{sec:background}
According to NIST Zero Trust (ZT) is defined as a “set of cybersecurity paradigms that move defences from static, network-based perimeters to focus on users, assets and resources”. 
According to NIST \cite{b6} there are three main approaches to implement ZTA including: enhanced identity governance; micro-segmentation; and network infrastructure using Software Defined Perimeters (SDP) \cite{moubayed2019software}. Over the past number of years, major industry players however have been converging on the use of SDP (Cisco, Intel, EMC) and SDP is now been used by cloud infrastructure providers (Amazon, Google) given its features of software defined abstractions, centralized management, automation and programmability features. 

The Cloud Security Alliance (CSA) proposed the concept of SDP as a method of deploying perimeters where needed replacing physical hardware appliances with logical components with the ability to protect networks in a dynamic manner. 
SDP architecture uses a controller to enforce that client applications/devices are first authenticated/authorized before creating encrypted connections in real time to the requested servers. SDP have a number of benefits including: hiding applications from unauthorised users; provide zero-visibility and zero-connectivity to all except authorized users and devices; enables operators to dynamically provision network perimeters; and works with user authentication systems \cite{moubayed2019software}. Despite the benefits, SDP also have a number of open challenges including issues around controller vulnerability and the overhead i.e. increased latency, in the control plane providing the features of authentication, access and encryption.   

To date, only a limited number of contributions have analysed Zero Trust implementations performance. Muji M. et al \cite{b8} analysed the performance of Zero Trust Architecture using micro-segmentation with results demonstrating that micro-segmentation added an average Round Trip Time (RTT) and jitter of 4 ms and 11 ms respectively without packet loss, demonstrating the viability of zero trust in a data centre environment. Surantha et al. \cite{b7} conducted a study to test the performance of network system security design using a zero-trust model. However, the work presented in \cite{b7} was focused more on the network security design features, rather on the performance overhead supporting cloud native applications in a hybrid cloud environment, which is the contribution of this proposed work. It is also important to note the work from Larsson L. et al. \cite{perf-istio}  which analysed the performance of Istio framework, which implements zero trust as part of it architectural solution against a native Kubernetes environment. However, the work focused on measuring throughput in a single cloud setup whereas the focus of this research will instead be on CPU, memory usage and latency of HTTP requests when Istio is deployed on a multi-cloud environment.

\section{Test-bed Architecture}
\label{sec:architecture}
A central element of the test bed architecture was the selection of an open source Zero Trust solution to perform the analysis. At present there are a number of  SDP/Zero Trust solutions available such as such as those provided by Project Calico \cite{calico}, Waverley Labs \cite{CSAZT} and the Istio Service Mesh \cite{istio}. In our architecture, the Istio Service Mesh was used, given its architectural design, separation of data and control planes, implementation of zero trust security principles, service based identity model and centralised view of security policies and adherence to defined policies \cite{sheikh2018modernize}. 

\begin{figure}[htbp]
\centerline{\includegraphics[width=90mm,scale=0.9]{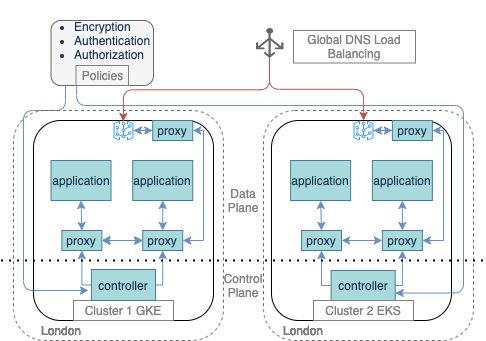}}
\caption{Architecture of scenario 2 – Zero Trust multi-cloud}
\label{fig4}
\vspace{-2mm}
\end{figure}

The architectural setup for the test bed is shown in Fig.~\ref{fig4}. The architectural setup involved two different cloud providers i.e. Kubernetes clusters, located in the same geographical region, London. Table \ref{tab:tab1} describes the hardware configuration for each cluster: Google Kubernetes Engine (GKE) and Elastic Kubernetes Service (EKS). 
In the architecture, a DNS load balancer was used to route traffic between the two Kubernetes clusters sitting in the two separate network domains i.e., Google Cloud and Amazon Web Services. The use of a load balancer was necessary in order to create a multi-cloud deployment to provide a benchmark for performance analysis (scenario 1). This benchmark was then compared to the same implementation with the added configuration of Istio Service Mesh to enable Zero Trust (scenario 2).


\begin{table}[htbp]
\vspace{-2mm}
\caption{Cluster Specification}
\vspace{-4mm}
\begin{center}
\resizebox{\columnwidth}{!}{%
\begin{tabular}{|c|c|c|c|c|c|}
\hline
\textbf{Cluster}&\textbf{Master Version}&\textbf{nodes}&\textbf{vCPU}&\textbf{Mem}&\textbf{image} \\
\cline{1-6} 

GKE&v1.18.16-gke.302&2&8&32Gb&E2-standard-4 \\
\hline
EKS&v1.18.9-eks-d1db3c&2&4&16Gb&T2.medium \\

\hline
\end{tabular}%
}
\label{tab:tab1}
\end{center}
\vspace{-2mm}
\end{table}

Leveraging this test bed, performance was benchmarked against two different configurations, with and without zero trust enforced. In each configuration, sequential HTTP requests were sent to a simple microservice application developed by Istio, which simply returned the pod name where the workload is running. The requests are sent in two modes: in cluster, from the Pod in the same cluster and name-space; cross cluster, from a Pod in another cluster. 

\section{Methodology}
\label{sec:methodology}
The methodology undertaken as part of this work involved a systems engineering approach with a focus on experimental evaluation. The experiments involved generating workloads within the clusters by running them in Kubernetes pods and measuring the resource usage at both a cluster and pod level. Resource usage Key Performance Indicators (KPIs) considered includ: Pod CPU usage and memory usage (while serving HTTP requests), and latency (HTTP requests response time and success rate). Further information on the resource and workload analysis are outlined below.

This method involves the analysis of CPU and memory system resources focusing on utilization to identify what percentage of resource usage is required while the system is responding to HTTP requests. 
Prometheus\cite{prometheus} was installed in each cluster to provide system visibility and access to resource metrics. Metrics are then extracted via a Grafana\cite{grafana} User Interface (UI) which uses Prometheus as a data source. 
This method is employed to analyse HTTP requests latency. Standard Linux utilities and cURL are employed to evaluate the latency by sending HTTP requests to a target workload running in a cluster. A bash script was developed to leverage those utilities and provide a test summary including percentiles, non-successful requests and the ratio of pods responding to the requests. The script can be executed on a local machine or containerized and run in any cluster as a Kubernetes pod.

\section{Results}
\label{sec:results}

As described in Section \ref{sec:architecture} a series of requests were sent as a test on various configurations of the testbed. Each test comprised 1000 sequential HTTP requests and all requests were successful (returned 200 HTTP response code). Table \ref{tab:tab2} represents a summary of results obtained for the tests in which GKE was responding to the HTTP request. The results presented indicate minimal impact on Cluster CPU and memory usage. However, the presence of the outlier value for the pod to service result with GKE using Istio regarding Cluster CPU max usage, suggests that additional factors may need to be examined such as the control plane resource consumption while the system is processing HTTP requests. It is important to note that percentage values presented are relative to cluster specification mentioned in Table \ref{tab:tab1}, where the GKE cluster has twice the number of cores and memory of EKS. 



\begin{table}[t!]
\vspace{-4mm}
\caption{GKE Response Performance analysis summary}
\vspace{-6mm}
\begin{center}
\resizebox{\columnwidth}{!}{%
\begin{tabular}{|c|c|c|c|c|c|}
\hline
\multicolumn{1}{|p{1.8cm}|}{\centering Request From}&
\multicolumn{1}{|p{1.8cm}|}{\centering Response From}&
\multicolumn{1}{|p{1.8cm}|}{\centering Pod CPU \\ max vCores}&
\multicolumn{1}{|p{1.8cm}|}{\centering Pod Memory \\ max MiB}&
\multicolumn{1}{|p{1.8cm}|}{\centering Cluster CPU max usage \%}&
\multicolumn{1}{|p{1.8cm}|}{\centering Cluster Memory max usage \%}\\
\hline
\multicolumn{6}{|c|}{\centering \textbf{Pod to Service with the same cluster}}\\
\hline
GKE&GKE&0.1&175&5.6&5.3 \\
\hline
GKE-ISTIO&GKE&0.9&85&17.8&5.6 \\
\hline
\multicolumn{6}{|c|}{\centering \textbf{Pod to Service across cluster via Gateway}}\\
\hline
EKS&GKE&1.2&183&19.1&5.4 \\
\hline
EKS-ISTIO&GKE&0.9&85&16.7&5.7 \\
\hline
\multicolumn{6}{|c|}{\centering \textbf{Pod to Service across cluster via DNS (Load Balanced)}}\\
\hline
GKE&GKE&1.1&183&19.3&5.5 \\
\hline
GKE-ISTIO&GKE&0.9&85&17.5&5.6 \\
\hline
EKS&GKE&0.7&183&12.7&5.5 \\
\hline
\end{tabular}%
}
\label{tab:tab2}
\end{center}
\vspace{-6mm}
\end{table}

\begin{table}[b!]
\vspace{-4mm}
\caption{EKS Response Performance analysis summary}
\vspace{-6mm}
\begin{center}
\resizebox{\columnwidth}{!}{%
\begin{tabular}{|c|c|c|c|c|c|}
\hline
\multicolumn{1}{|p{1.8cm}|}{\centering Request From}&
\multicolumn{1}{|p{1.8cm}|}{\centering Response From}&
\multicolumn{1}{|p{1.8cm}|}{\centering Pod CPU \\ max vCores}&
\multicolumn{1}{|p{1.8cm}|}{\centering Pod Memory \\max MiB}&
\multicolumn{1}{|p{1.8cm}|}{\centering Cluster CPU max usage \%}&
\multicolumn{1}{|p{1.8cm}|}{\centering Cluster Memory max usage \%}\\
\hline
\multicolumn{6}{|c|}{\centering \textbf{Pod to Service with the same cluster}}\\
\hline
EKS&EKS&1.1&230&33.3&17.2 \\
\hline
EKS-ISTIO&EKS&0.9&115&30.1&18.7 \\
\hline
\multicolumn{6}{|c|}{\centering \textbf{Pod to Service across cluster via Gateway}}\\
\hline
GKE&EKS&1.0&215&31.3&17.6 \\
\hline
GKE-ISTIO&EKS&0.9&115&26.9&19.0 \\
\hline
\multicolumn{6}{|c|}{\centering \textbf{Pod to Service across cluster via DNS (Load Balanced)}}\\
\hline
GKE&EKS&1&215&27.3&17.7 \\
\hline
GKE-ISTIO&EKS&0.8&115&25.7&19.0 \\
\hline
EKS&EKS&0.6&215&21.4&17.6 \\
\hline
EKS-ISTIO&EKS&0.9&115&28.9&19.0 \\
\hline
\end{tabular}%
}
\label{tab:tab3}
\end{center}
\end{table}

Table \ref{tab:tab3} represents a summary of results obtained for the tests in which EKS was responding to the HTTP requests. For the tests presented cluster CPU and memory max usage do not significantly vary. In both Table \ref{tab:tab2} and Table \ref{tab:tab3} pod memory usage with the Istio service mesh is reduced by ~50\%.  This reduction is most likely a result of offloading local process from the pod to the control plane.
 
\begin{figure}[htbp]
\centerline{\includegraphics[width=90mm,scale=0.9]{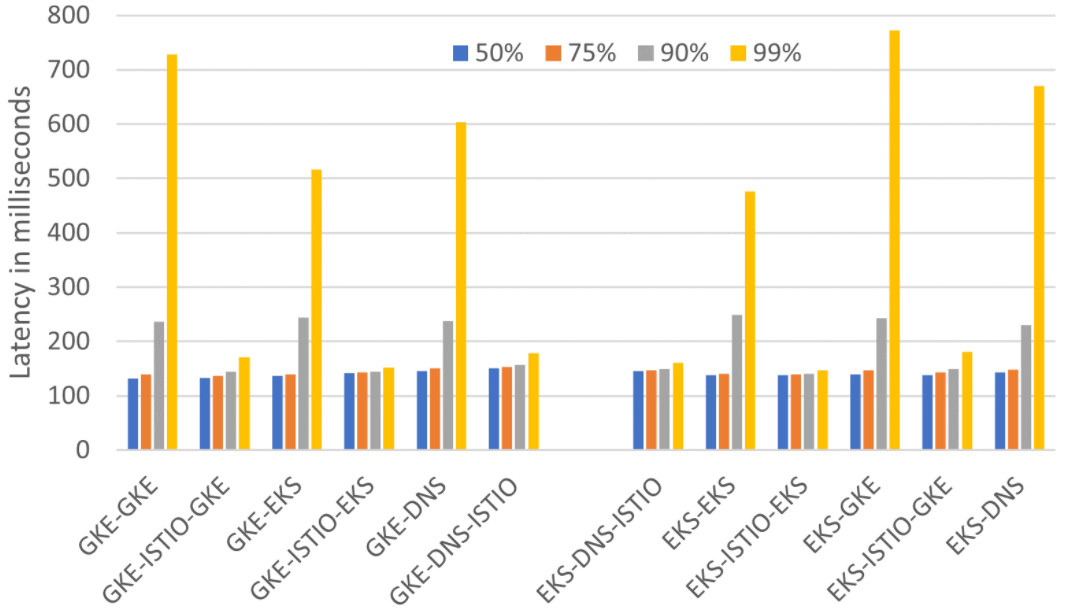}}
\vspace{-2mm}
\caption{Latency Test results}
\label{fig3}
\vspace{-4mm}
\end{figure}

Fig.~\ref{fig3}, represent an interesting finding in the analysis of the request group latency. For each test run the latency in milliseconds is displayed on the vertical axis for 50\%, 75\%, 90\%, and 99\% of responses returned. The workload analysis in Fig.~\ref{fig3} shows higher latency for simple multi-cloud deployment, particularly for 75\% and above. This indicates that Istio Ingress gateway resource handles networking requests in a more efficient manner than the basic Load Balancer service from Kubernetes. 

The preliminary results presented in this paper show no evident performance penalty at the data-plane level when enabling Zero Trust with a side-car proxy implementation in a multi-cloud architecture using Istio service mesh. The experiments were appropriately conducted with respect to the objectives of the study in terms of configurations of multiple PCP environments, service mesh and policy enforcement to include encryption, authentication and authorization. However, the number of tests and test scenarios should improve to provide statistical significance in terms of resource analysis. 



\section{Discussion and Future work}
\label{sec:discussion}
As part of the work presented, a performance analysis of the data plane under load and the impact on the control plane was measured by analysing 
cluster and pod resource consumption.
Test results demonstrate that Istio has reduced latency variability in responding to sequential HTTP requests. However, system resources and particularly overall CPU and memory can increase based on service mesh configuration and the cloud environment. In order to explore the impact of this further, a more comprehensive set of tests and scenarios are required. In addition, for future tests we will allocate images with similar specifications for both managed clusters for memory and CPU cores, and will also include a more detailed performance analysis of the control plane under load further evaluating Zero Trust.


\bibliographystyle{IEEEtran}
\bibliography{refs}

\end{document}